\documentclass[conference]{IEEEtran}
\IEEEoverridecommandlockouts
\usepackage{cite}
\usepackage{amsmath,amssymb,amsfonts}
\usepackage{mathtools}
\usepackage{algorithmic}
\usepackage{graphicx}
\usepackage{caption}
\usepackage{subcaption}
\usepackage{wrapfig}
\usepackage{textcomp}
\usepackage{xcolor}
\usepackage{url} 
\usepackage{soul,color}
\usepackage{adjustbox}
\usepackage{comment}
\usepackage[hidelinks]{hyperref}
\usepackage{orcidlink}
\usepackage{multirow} 
\usepackage{enumitem}
\usepackage{booktabs}

\usepackage[capitalize]{cleveref}
\crefname{section}{Sec.}{Secs.}
\Crefname{section}{Section}{Sections}
\Crefname{table}{Table}{Tables}
\crefname{table}{Tab.}{Tabs.}

\def \ie {\emph{i.e.},}
\def \eg {\emph{e.g.},}

\hypersetup{draft}
\begin{document}

\title{RI-PIENO -- Revised and Improved Petrol-Filling Itinerary Estimation aNd Optimization}

\author{\IEEEauthorblockN{Marco Savarese$^1$\orcidlink{0009-0004-9722-0110}, Antonio De Blasi$^1$
\orcidlink{0009-0004-3899-8424}, Carmine Zaccagnino$^{1, 2}$\orcidlink{0009-0001-2069-6776}, Giacomo Salici$^{1, 2}$\orcidlink{0009-0004-4953-6348},\\Silvia Cascianelli$^2$\orcidlink{0000-0001-7885-6050}, Roberto Vezzani$^2$\orcidlink{0000-0002-1046-6870}, Carlo Augusto Grazia$^2$\orcidlink{0000-0003-0534-995X}}
\IEEEauthorblockA{
\textit{University of Modena and Reggio Emilia}\\
$^1$\{name.surname\}@pieno.dev, $^2$\{name.surname\}@unimore.it
}
}

\maketitle

\begin{abstract}
Efficient energy provisioning is a fundamental requirement for modern transportation systems, making refueling path optimization a critical challenge. Existing solutions often focus either on inter-vehicle communication or intra-vehicle monitoring, leveraging Intelligent Transportation Systems, Digital Twins, and Software-Defined Internet of Vehicles with Cloud/Fog/Edge infrastructures. However, integrated frameworks that adapt dynamically to driver mobility patterns are still underdeveloped.
Building on our previous PIENO framework, we present RI-PIENO (Revised and Improved Petrol-filling Itinerary Estimation aNd Optimization), a system that combines intra-vehicle sensor data with external geospatial and fuel price information, processed via IoT-enabled Cloud/Fog services. RI-PIENO models refueling as a dynamic, time-evolving directed acyclic graph that reflects both habitual daily trips and real-time vehicular inputs, transforming the system from a static recommendation tool into a continuously adaptive decision engine.
We validate RI-PIENO in a daily-commute use case through realistic multi-driver, multi-week simulations, showing that it achieves significant cost savings and more efficient routing compared to previous approaches. The framework is designed to leverage emerging roadside infrastructure and V2X communication, supporting scalable deployment within next-generation IoT and vehicular networking ecosystems.
\end{abstract}


\section{Introduction}
\label{sec:introduction}
Wherever transportation of people or goods is required, the supply of energy becomes a fundamental constraint. This has motivated decades of research on the optimization of refueling paths, with approaches ranging from network-level routing optimization to AI-assisted driving strategies. These include the gas station problem \cite{khuller2011tofill}, which has also been tackled with Reinforcement Learning\cite{Ottoni2022reinforcement}.
On the networking side, Software-Defined Internet of Vehicles has been applied to QoS-aware routing~\cite{SDIV_Routing}, RSU cloud services~\cite{SDIV_RSU}, and heterogeneous multi-radio access management~\cite{SDN_Access1}, while delay-tolerant schemes have been explored in cloud–fog vehicular systems~\cite{SDN_Access2}. Other contributions address time-sensitive in-vehicle networks~\cite{SDN_Access3} and cooperative paradigms such as platooning~\cite{tits10,tits3}. Meanwhile, AI-based methods have been proposed to optimize energy consumption by learning from driver behavior~\cite{tits7}. Despite these advances, most works treat either inter-vehicle communication or intra-vehicle monitoring in isolation, and only rarely provide an integrated solution that is both practically deployable and adaptable across diverse mobility scenarios.

In our earlier work, \textit{Petrol-Filling Itinerary Estimation aNd Optimization} (PIENO)~\cite{savarese2024pieno}, we presented a proof-of-concept framework combining intra-vehicle data (\eg~fuel level through CAN Bus and OBD-II~\cite{OBDII,MCP2515,SAE-J1979}) with external data on fuel prices and petrol station locations, processed via cloud services. PIENO demonstrated how Intelligent Transportation Systems and Digital Twin concepts~\cite{digitaltwin} can be combined to deliver personalized, eco-aware refueling recommendations. However, PIENO was conceived as a static framework: it focused on optimizing the next refueling event, employed simplified behavioral heuristics, and lacked mechanisms to adapt dynamically to recurring travel routines or longer-term usage contexts.

In this paper, we present a revised and improved version of PIENO, dubbed RI-PIENO, which addresses the aforementioned limitations. Specifically, RI-PIENO extends the PIENO framework by introducing: 
\begin{itemize}
    \item Temporal modeling of a driver’s habitual daily trips; 
    \item Integration of machine learning predictions of user habits into Operations Research-based optimization; 
    \item Representation of the refueling problem as a directed acyclic graph that evolves in time based on real-time sensor and geospatial inputs.
\end{itemize}
The combination of the original PIENO features and these novel contributions transforms the system from a one-off decision support tool into a continuously adaptive engine for refueling path optimization. 
Furthermore, the framework is designed to leverage emerging roadside infrastructure and V2X communication, paving the way for scalable, large-scale deployment scenarios.

We present RI-PIENO in the use-case of personalized daily commutes and show how to mathematically model this scenario as a directed acyclic graph, built over time based on real data from the car's sensors and geospatial information. Note that, although beyond the scope of this paper, the same formalism we present is straightforward to apply to the use-case of single-day special journeys, which can be modeled as a simpler, shorter graph. 
We validate RI-PIENO through realistic simulations involving multiple drivers and paths over several weeks. Our results demonstrate that RI-PIENO outperforms both PIENO and baseline strategies, both in terms of cost and time saved, and also has a direct impact on reducing the user's emissions by optimizing their refueling itinerary.

\begin{figure}[t]
    \centering
    \includegraphics[width=\linewidth]{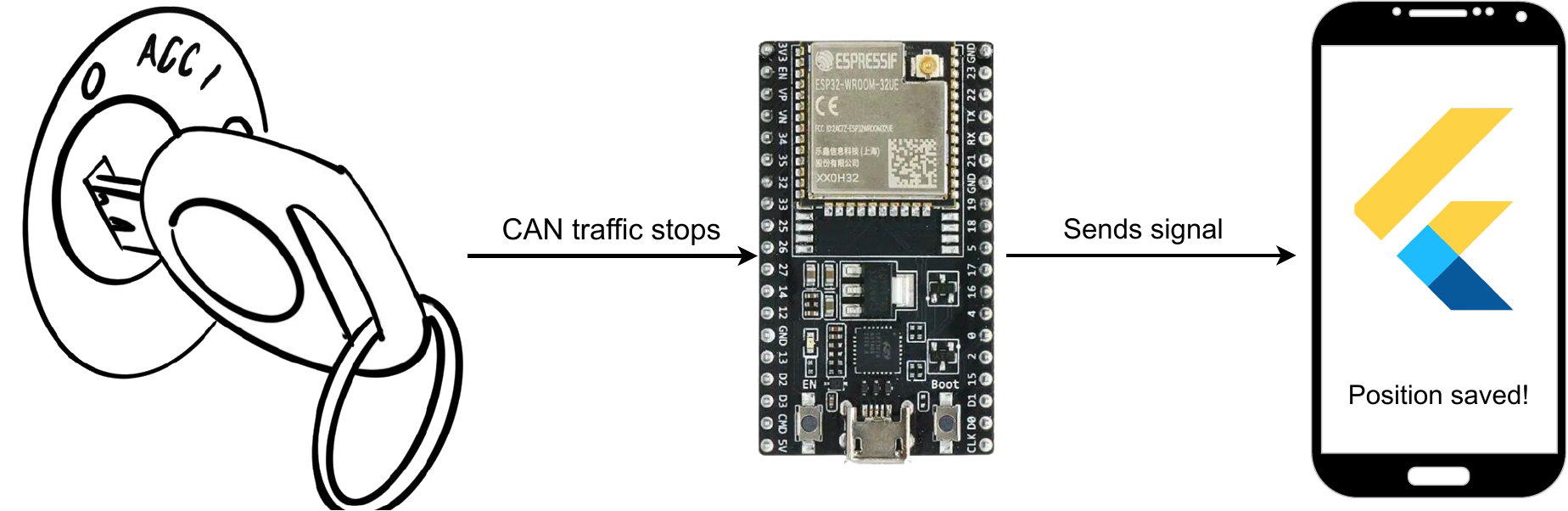}
    \caption{Communication between RI-PIENO components.}
    \label{fig:comm}
\end{figure}

\section{Method}
\label{sec:method}

In this section, we present the main modules of the RI-PIENO framework, specifically designed to overcome the main limitation of PIENO~\cite{savarese2024pieno}, namely its lack of integration of user mobility patterns into the optimization process:
\begin{enumerate}[label=\emph{\textbf{\Alph*.}}, itemsep=0.2em]
    \item \emph{\textbf{Daily Trip Graph}}: constructs a directed graph for each day of the week based on the user’s identified \emph{Points of Interest} (POIs). Each graph captures the principal daily movements and recurrent trip sequences.
    \item \emph{\textbf{Weekly Mileage Prediction}}: applies a machine learning approach to forecast the number of kilometers the user is expected to drive in the upcoming week. The prediction exploits historical trip data and an ensemble strategy capturing both non-linear patterns and overall trends.
    \item \emph{\textbf{Fuel Stop Optimization}}: leverages the information encoded in the daily trip graph together with an open-source routing engine to compute the optimal one-stop refueling route. Unlike PIENO, the stop is selected in coherence with the user’s habitual trajectories.
\end{enumerate}

\subsection{Daily Trip Graph.}\label{ssec:graph} 
The \emph{Daily Trip Graph} extraction module is in charge of modeling user-specific mobility patterns and information about the usual routes, which will be used by the \textit{Fuel Stop Optimization} module. Specifically, this module processes raw location data, identifies the POIs, and reconstructs daily flow sequences in the form of directed graphs, whose nodes are the user's POIs and edges represent the usual route taken by the user to move from one POI to another. The graph is built on the basis of data collected from the vehicle during an observation period (in this work, we consider at least two weeks).
The rationale behind this approach is to capture the typical flow of movements between usual destinations on a per-day basis, allowing the system to distinguish between regular mobility patterns (\eg~\emph{home} $\rightarrow$ \emph{work} $\rightarrow$ \emph{gym}) and incidental stops, thereby enriching the system with personalized mobility knowledge.  
The module is organized into two main components:
\begin{enumerate}[label=\emph{\textbf{\arabic*)}}, itemsep=0.2em]
    \item \emph{\textbf{POI Creator}}, responsible for clustering raw GPS coordinates into POIs based on spatial and temporal criteria;
    \item \emph{\textbf{Flow Generator}}, responsible for reconstructing daily trip flows by linking POIs into chronological sequences.
\end{enumerate}
Both the POI Creator and the Flow Generator rely on a common base data structure that integrates information from the CAN Bus and the ESP32 ~\cite{ESP32}, part of the CAM of PIENO~\cite{savarese2024pieno}, and is stored in the PIENO mobile application~\cite{savarese2024pieno}. The specifics of this process will be described in \Cref{ssec:poicreat} while their role and their output are detailed in the remainder of this subsection.

\subsubsection{POI Creator}\label{ssec:poicreat}

\begin{figure}[t]
    \centering
    \includegraphics[width=.8\linewidth]{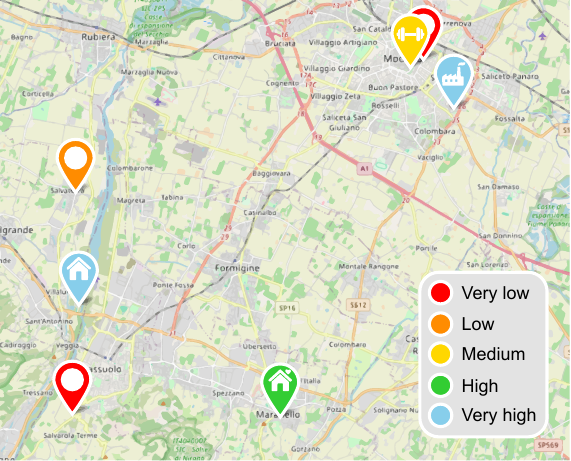}
    \caption{Geographical map of all the clustered stops that a user has performed during the observation period, colored based on their frequency of visit that week. Note that only stops visited with at least Medium frequency are considered as POIs.}
    \label{fig:pois}
\end{figure}

To build the daily trip graph, we first detect the vehicle stops through the CAN Bus. Specifically, when the car comes to a halt, the message flow on the CAN Bus is interrupted. Since these two events are strongly correlated, the absence of messages can be used as a proxy for vehicle stop detection. Once a stop is detected, the ESP32 sends a signal to the PIENO app, and the corresponding GPS position is stored and enriched with contextual metadata such as time, date and timestamp. This data flow is depicted in~\cref{fig:comm}. Note that, to prevent storing multiple stops all referring to the same location, we cluster new stops with previously recorded ones if they fall within a predefined radius (set to 100m in this study). This design choice also models the scenario in which the user parks in different spots close to their actual destination. 

Each stop is a candidate POI node and is associated with a rich set of metadata:
\begin{itemize}
    \item \textbf{Identifier:} Unique node label (\eg~ \texttt{STOP\_001}).
    \item \textbf{Coordinates:} Latitude and longitude of the centroid of the cluster to which nearby stops have been associated. These coordinates are obtained by combining those recorded by the PIENO app from the smartphone's GPS at the stops in the cluster.
    \item \textbf{Visit frequency:} Total number of visits to that location registered during the observation period, disaggregated into weekdays and weekends to enable refined analyses (even if beyond the scope of this work).
    \item \textbf{Frequency category:} Automatically assigned label based on the average number of visits to that node in a week ($v$). Specifically, the categories are \texttt{VERY\_LOW} ($v\leq$1), \texttt{LOW} (1${<}v{\leq}$2), \texttt{MEDIUM} (2${<}v{\leq}$4), \texttt{HIGH} (4${<}v{<}$10), and \texttt{VERY\_HIGH} ($v{\geq}$10).  
    \item \textbf{Days visited:} Set of days in which the POI has been visited during the observation period (\eg~\texttt{Mon$|$Tue$|$Wed}). These are reconstructed from the date metadata registered by the PIENO app at corresponding halt events.
\end{itemize}
A stop is then considered a user POI if it has been assigned one of the Frequency category labels from a pre-defined set. In this work, we consider \texttt{VERY\_HIGH}, \texttt{HIGH}, and \texttt{MEDIUM} as acceptable labels (see~\cref{fig:pois}).
This structured information allows the system to distinguish highly relevant POIs (\eg~home, workplace) from less significant ones (\eg~occasional shops).

\subsubsection{Flow Generator}
\begin{figure}[t]
    \centering
    \includegraphics[width=\linewidth]{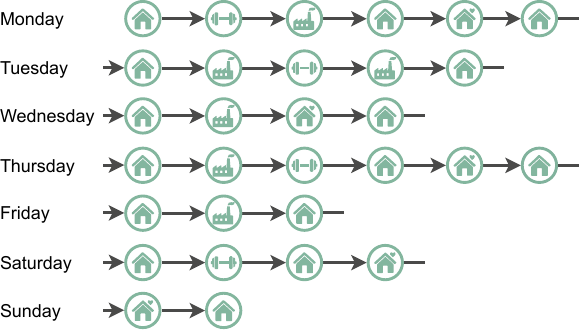}
    \caption{Daily Trip Graph obtained after the user's observation period, which reflects their typical weekly trips (\ie~between ordered POIs) day by day.}
    \label{fig:flow}
\end{figure}

The edges of the Daily Trip Graph represent the directed transitions between POIs observed during a single day. 
These are obtained by sorting the POI nodes based on the timestamp metadata registered by the PIENO app at corresponding halt events, and then tracing a directed edge from $\mathrm{POI}_i$ to $\mathrm{POI}_j$ between consecutive POIs (see~\cref{fig:flow}).

For each transition, we store metadata associated with its destination node, with the implicit assumption that the previous entry represents the source. The stored information for each edge includes:
\begin{itemize}
    \item \textbf{Day Identifier:} Tag that enables the optimizer (described in \Cref{ssec:FSO}) to restrict the analysis to the specific day under consideration.
    \item \textbf{Sequence index:} Integer that encodes the chronological order of the POI within the daily trajectory. This value is derived directly from the ordering of the stop events in the data structure of the application.
    \item \textbf{Coordinates:} Latitude and Longitude of the destination POI node.
\end{itemize}
\vspace{0.25em}
From a practical point of view, the output of the Daily Trip Graph extraction module consists of two structured CSV files, one for the POI Nodes and one for the daily transitions (\ie~the edges), containing their metadata. This modular design is straightforward to adapt and apply to novel scenarios and different applications (\eg~it is easy to change the definition of POI Nodes, or to generate alternative transition hypotheses or routes for the same POIs). Therefore, although the scope of this work is fuel-stop recommendation, it is worth noting that the devised daily trip graph provides a compact yet expressive representation of mobility patterns that can also support tasks like next-destination prediction, flow analysis, or personalized route optimization.

\subsection{Weekly Mileage Prediction.} \label{sec:method_mileage_pred} 
With the Daily Trip Graph, we obtain a picture of the consistent trips and the usual locations visited by a user over the week. Nonetheless, the user likely makes occasional trips to other locations. These are basically non-POI nodes. Therefore, in addition to their usual paths (and associated traveled distance, which can be obtained almost deterministically), the user will also travel to non-POI nodes every week. However, since these unusual locations are likely different from week to week, the path and traveled distance to reach them, which has an impact on the user's fuel consumption, cannot be known a priori. To alleviate this issue, we devise a machine learning-based approach to estimate the mileage a user is likely to drive in the following week. Specifically, we employ a Random Forest~\cite{Breiman_2001_rf} Regressor trained on historical trip data. The hyperparameters of the model are set to 150 estimators and a maximum depth of 6 after empirical analysis.

The predictor variables (features) capture both temporal patterns (such as the \textbf{day of the week} and the \textbf{month of the year}) and behavioral dynamics of travel (including \textbf{short-term lags} and \textbf{rolling averages of distance traveled}). In addition, when available, we include indicators of trip intensity and performance, such as the \textbf{number of trips}, and measures of \textbf{average and maximum speed}. Note that, to ensure comparability and improve model stability, we standardize all features prior to training. 

We train the model on data collected during a variable number of weeks and validate it on an unseen (recent) week. 
To measure the model's performance, we use the \textit{mean absolute error (MAE)} computed between all the daily predictions, \ie
$$    \text{MAE} = \frac{1}{N} \sum_{n=1}^{N} \left| y_n - \hat{y}_n \right|, $$
the \textit{weekly absolute error}
$$    E_{\text{week}} = \left| \sum_{n \in W} y_n - \sum_{n \in W} \hat{y}_n \right|, $$
and the \textit{weekly percentage error}
$$    E_{\text{week\%}} = \frac{E_{\text{week}}}{\sum_{n \in W} y_n} \times 100, $$
where $y_n$ is the actual daily distance, $\hat{y}_n$ is the predicted daily distance, $W$ denotes the set of days in the week, and $N$ is the total number of days evaluated. 

Note that this data-driven predictive approach works best for users with consistent habits, which is not always the case. Therefore, RI-PIENO applies the model to predict the user's mileage for the following week only if the validation is successful (\ie~each prediction error is lower than the predefined threshold set as: $\text{MAE}\leq $2.5, $E_{\text{week}}\leq $5.7, and $E_{\text{week\%}} \leq $21.3, in this work). 

\subsection{Fuel Stop Optimization.}\label{ssec:FSO} 
Similar to what done in PIENO~\cite{savarese2024pieno}, we exploit the weekly forecast of the fuel costs in the user's area to select the most convenient day of the week for refilling. This directly translates into a portion of the Daily Trip Graph described in~\Cref{ssec:graph}. For that day, we compute the expected traveled distance between the POIs, $d_n$ and use the estimated weekly mileage $\hat{y}_n$ (if deemed sufficiently reliable) to obtain an estimation of the extra distance traveled to reach unusual locations as:
$$
\delta_n = \hat{y}_n - d_n.
$$
In the following, we drop the suffix $n$ for brevity.

Given graph for the selected day, we find potential fuel stops along the user's path, which we call \emph{Fuel Station Nodes} (PS Nodes). To this end, we first compute the optimal routing between the ordered POIs using the open-source GraphHopper\footnote{\url{https://graphhopper.com/}} routing engine based on OpenStreetMap\footnote{\url{https://www.openstreetmap.org}} data (nonetheless, note that RI-PIENO is independent on the routing engine and can be used directly also with other systems, \eg~OSRM). Then, as in PIENO~\cite{savarese2024pieno}, we leverage government-provided fuel station data to obtain an exhaustive list of fuel stations in the user's area, for which we also collect location coordinates and fuel prices. 
Finally, we filter the fuel stations by using their latitude and longitude and only keeping fuel stations that fall within a range of 2 km from the optimal route between the POIs (for computational reasons, but this radius can be enlarged). 
We refer to this list of candidate stops as $PS$.
These PS nodes are \textit{candidate refueling stops}, among which the best one will be selected by exploiting our devised optimization algorithm. 
For each candidate stop, we compute (again by exploiting GraphHopper) all possible one-stop routes starting from the closest, preceding POI node along the Daily Trip Graph. The output we get from GraphHopper includes time $t_k$ and distance $l_k$ to travel to each fuel station $PS_k$ and reach the remaining POIs for the day, which can entail a short detour if the fuel station is not exactly along the user's path. The distance value is corrected to account for weekly mileage variations, as $\hat{l}_k = l_k+\delta$.  
From these values, we obtain the fuel and time cost associated with the PS nodes, which are the input to our algorithm. The details are given below.

\noindent\textbf{Fuel Cost.}
The fuel cost associated with the $k$-th PS node is computed as
$$
C_k =  (f_{full}-f_0 + r\cdot \hat{l}_k)\cdot c_k,
$$
where $f_{full}$ is the tank capacity of the vehicle, $f_0$ is the fuel level taken from the CAN Bus at the start of the trip, $r$ is an estimate of the fuel consumption rate of the vehicle, and the $C_k$ is the cost of the fuel at the fuel station $PS_k$, derived from the government data.

\noindent\textbf{Time Cost.}
We determine the time cost associated with stopping at a given fuel station $PS_k$ as the sum of the driving time estimated by GraphHopper $t_k$ and a fixed estimation of refueling time $\Delta$:
$$
T_k = t_k + \Delta.
$$
Note that, in this work, we consider $\Delta$ as a constant with the same value for all the fuel stations (set to 5 minutes, in this work). Therefore, it will not have an impact on the selection of the best one. However, it is worth noting that this mathematical model allows for straightforward future extension in which $\Delta$ can vary depending on additional information coming from the fuel station (\eg~the number of vehicles currently queuing at the station) or other users connected via the PIENO app.\vspace{0.3em}

Given all the candidate fuel stops and their associated information, we first distinguish between reachable and unreachable stops. Unreachable stops are those that cannot be reached with the current level of fuel $f_0$ due to their distance $\hat{l}_k$ and the user's fuel consumption rate $r$. The optimization cost of these stops is set to infinity. Then, we find the index $k$ of the optimal fuel station among the candidate stops as the one that minimizes the overall cost function:
$$
\mathcal{L}_k = \begin{cases}
    K_1\cdot C_i+K_2\cdot T_i & \text{if  }  r\cdot \hat{l}_k \le f_0 \\
    \infty & \text{if }  r\cdot \hat{l}_k > f_0,
\end{cases}
$$
where the two constants $K_1$ and $K_2$ are used to model the user's preference to save cost or time.
Specifically, by tweaking these two constants, we hypothesize three possible \textit{modes} of operation of the system:
\begin{itemize}
    \item \textbf{Fuel-sensitive}, if the user is particularly attentive to saving on fuel cost, and does not mind small detours to reach the optimal fuel destination. In our experiments, we set $K_1 = 1$ and $K_2 = 0$ to model this scenario.
    \item \textbf{Time-sensitive}, if the user is particularly attentive to saving time, and does not mind paying more for fuel. In our experiments, we set $K_1 = 1$ and $K_2 = 10$ to model this scenario.
    \item \textbf{Balanced}, if the user aims at balancing the two components. This is the default mode of the RI-PIENO system presented in this paper, for which we set the constants as $K_1 = 1$ and $K_2 = 1$.
\end{itemize}

\section{Experiments and Results}
\label{sec:results}
In this section, we validate our proposed approach and its components on a public dataset and on a set of recordings collected for this work.

\subsection{Weekly Mileage Prediction.}\label{sec:results_mileage_pred} 
To evaluate the proposed machine learning-based approach for the weekly mileage prediction on an independent set of data, we apply it to the VED dataset~\cite{oh2022ved}, a large-scale collection of operational data from hundreds of vehicles in real-world conditions. This dataset contains measurements of energy consumption and dynamic movement information, including GPS coordinates and speed, sampled every 1-3s seconds. To use the VED dataset in our experiments, we need to obtain a ground truth estimate for the daily distance traveled by each driver in the dataset. To this end, we aggregate the raw measurements available into total daily usage per vehicle. Specifically, we multiply the average speed recorded by the elapsed time between sampled positions. Note that we decide not to use a GPS-dependent approach (\eg~calculating the geodesic distance between consecutive GPS samples based on their coordinates) since GPS signals may not always be available in real-world scenarios. Nonetheless, we have empirically found that our approach and a GPS-based one yield nearly identical results (0.166 km of median difference on their estimations).

For model evaluation, we perform time-series cross-validation using a sliding-window approach, where the training set consists of samples collected during a fixed number of historical weeks and the prediction horizon is the following week. The size of the historical training window is varied to examine its effect on the predictive performance. At each iteration, both daily and weekly forecasts are computed, and the results are recorded for aggregated statistical analysis across all validation folds.
~\Cref{tab:rf_results} presents the results of the Random Forest model when trained on data from different window sizes. The reported values are averaged over all 45 vehicles that traveled more than 1000 km during the one-year period covered by the VED dataset. The results demonstrate that with sufficiently long observation periods, the weekly mileage estimation is increasingly precise. Moreover, the relatively high standard deviations across predictions indicate that some travel profiles are more predictable than others (as argued in~\Cref{sec:method_mileage_pred}). By taking into account these results on the independent VED dataset, we set the acceptability thresholds for integrating the output of the Weekly Mileage Prediction module into the  RI-PIENO optimization. 
Specifically, we use the 25$^{\text{th}}$ percentile of the results obtained by training on 6 weeks worth of data.

\begin{table}[t]
\centering
\resizebox{\linewidth}{!}{%
\begin{tabular}{lccc}
\toprule
\textbf{Obs. Period} & \textbf{MAE} & \textbf{$E_{\text{week}}$} & \textbf{$E_{\text{week\%}}$} \\
\midrule
4 weeks & 4.27 $\pm$ 1.62 & 10.09 $\pm$ 4.50 & 42.0 $\pm$ 28.0 \\
6 weeks & 3.86 $\pm$ 1.54 & ~9.04 $\pm$ 4.08 & 38.0 $\pm$ 25.1 \\
8 weeks & 3.59 $\pm$ 1.47 & ~8.52 $\pm$ 3.97 & 34.2 $\pm$ 21.3 \\
\bottomrule
\end{tabular}}
\caption{Random Forest performance on the same week, when trained on data collected during preceding observation periods (Obs. Period) of different length.}
\label{tab:rf_results}
\end{table}

\subsection{Comparison with PIENO.} 
To assess the benefit of the optimization strategy in RI-PIENO, we consider 15 recordings obtained from 3 different users on different paths and periods, and compare it against  PIENO and a baseline that entails simply stopping at the fuel station that is the closest one to the current user's location, regardless of the fuel cost at that station.

Note that, for fair comparison between RI-PIENO and the competitors, we consider the visited locations and traversed paths from the users' recordings and disregard their actual refueling stops. From these sets of data, we simulate what would have happened if the users were following the refilling advice given by the considered approaches, when departing from one of their POIs. In this way, we remove any variability due to daily external factors (\eg~traffic, driver conditions). 

\begin{figure*}
    \centering
    \setlength{\tabcolsep}{4pt}
    \begin{tabular}{ccc}
    \includegraphics[width=.3175\linewidth]{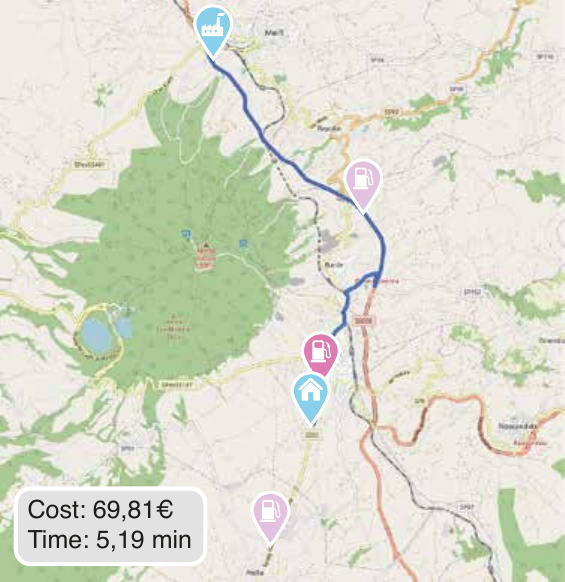} &
    \includegraphics[width=.3175\linewidth]{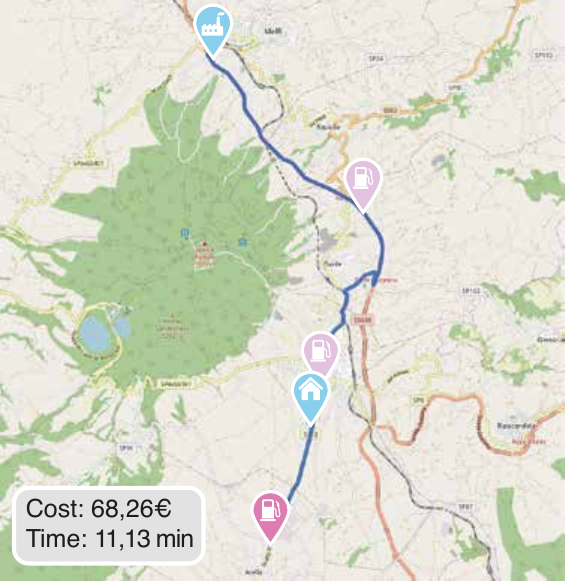} &
    \includegraphics[width=.3175\linewidth]{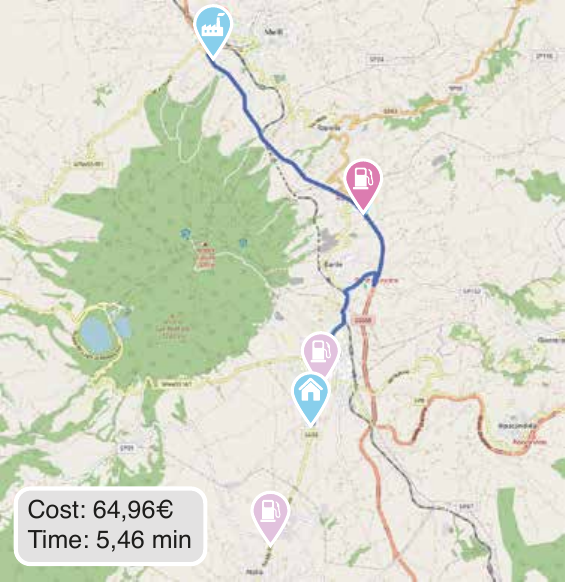} 
    \end{tabular}
    \caption{User path from a POI (home) to another POI (workplace) and Fuel Station stop (dark pink) proposed by the Baseline (left), PIENO (center), and RI-PIENO (right), with relative cost and time. The fuel stations in the area are in light pink.}
    \label{fig:comparison_maps}
\end{figure*}

For each path in the recordings set, we take the cost and time of refilling obtained with these approaches and average the results over all the recordings in the set. We report the obtained mean and standard deviation in ~\Cref{tab:comparison_results}. 
As it can be observed, RI-PIENO consistently leads to cost and time savings. This performance improvement can be attributed to the fact that, different from PIENO, RI-PIENO takes into account the path that the user will likely take from the current POI (thanks to the Daily Trip Graph that models their habits). This is also showcased in~\cref{fig:comparison_maps}, where we report the fuel station stop suggested by the Baseline, PIENO, and RI-PIENO to a user departing from a POI and directed to another one.

\begin{table}[t]
\centering
\resizebox{.7\linewidth}{!}{%
\begin{tabular}{l cc}
\toprule
 & \textbf{Cost [€]} & \textbf{Time [min]} \\
\midrule
Baseline & 44.53 $\pm$ 2.20 & 6.61 $\pm$ 1.01 \\
PIENO    & 43.24 $\pm$ 2.09 & 7.10 $\pm$ 2.23 \\
RI-PIENO & 43.01 $\pm$ 2.13 & 4.98 $\pm$ 0.80 \\
\bottomrule
\end{tabular}}
\caption{Comparison between our proposed RI-PIENO, and simpler approaches that do not take into account the user's behavior (PIENO) and the fuel cost information (Baseline).}
\label{tab:comparison_results}
\end{table}

\begin{table}[t]
\centering
\resizebox{.7\linewidth}{!}{%
\begin{tabular}{cc cc}
\toprule
\textbf{$K_1$} & \textbf{$K_2$} & \textbf{Cost [€]} & \textbf{Time [min]} \\
\midrule
1 & 0 & 42.64 $\pm$ 1.77 & 7.65 $\pm$ 4.46\\
10 & 1 & 42.72 $\pm$ 1.84 & 5.50 $\pm$ 0.49 \\
1 & 1 & 43.01 $\pm$ 2.13 & 4.98 $\pm$ 0.80 \\
1 & 10 & 43.67 $\pm$ 1.98 & 4.87 $\pm$ 0.76\\
0 & 1 & 44.38 $\pm$ 2.44 & 4.85 $\pm$ 0.49\\
\bottomrule
\end{tabular}}
\caption{Effect of the optimization constants on the RI-PIENO solution. The first two rows reflect the \textit{Fuel-sensitive} mode, the third the \textit{Balanced} mode, and the last two the \textit{Time-sensitive} mode.}
\label{tab:modes_results}
\end{table}

\subsection{Operation Modes.} 
The most relevant hyperparameters of RI-PIENO are the optimization constants $K_1$ and $K_2$, that model the importance to be given to saving fuel cost and time, respectively. Recall that different values given to the constants lead to different operation modes: Balanced ($K_1=K_2$), which is the default setting, Fuel-sensitive ($K_1>K_2$), and Time-sensitive ($K_1<K_2$), which can be made more or less aggressive by changing the difference between the two constants. In~\Cref{tab:modes_results}, we report the results obtained when changing their values in terms of saved cost and time. The results confirm the flexibility of the RI-PIENO approach: in the \textbf{Fuel-sensitive mode} (first two rows) we maximize the cost savings (as low as 42.64~€) at the expense of higher travel time (up to 7.65~min). On the contrary, in the \textbf{Time-sensitive mode} (last two rows) refueling time is heavily reduced (around 4.85~min) at the expense of reducing cost efficiency (up to 44.38~€). Finally, the \textbf{Balanced mode} (third row) offers an intermediate solution that overcomes both the Baseline and PIENO strategies.

\section{Conclusions}
\label{sec:conclusions}

In this paper, we have presented \emph{RI-PIENO}, an improved and extended version of the PIENO framework for refueling path optimization. By integrating temporal modeling of daily trips, a time-evolving directed acyclic graph representation, and an Operations Research-based optimization module, RI-PIENO transforms the refueling recommendation process from a static decision tool into a continuously adaptive system.  
Another key strength of RI-PIENO lies in its flexibility. In fact, the system design makes use of several parameters, such as the fuel consumption rate $r$, the refueling time $\Delta$, and the user preference weights $K_1$ and $K_2$ that balance cost and time. In this work, we have set these parameters to fixed, pre-defined values. Nonetheless, in future implementations, these parameters can be either refined through long-term observation of user behavior (\eg~empirical estimations of $r$ from CAN Bus data) or explicitly set by the user through the application interface. This makes the framework straightforwardly extendable to a wide variety of driving styles, vehicle types, and user preferences.  
We have validated the efficacy of RI-PIENO through simulations based on real-world data collected from multiple users over several days. The results demonstrate significant savings and improved routing compared to both PIENO and baseline strategies.

\section*{Acknowledgment}
This work was partially supported by MOST – Sustainable Mobility National Research Center and received funding from the European Union Next-GenerationEU (Piano Nazionale di Ripresa e Resilienza (PNRR) – Missione 4 Componente 2, Investimento 1.4 – D.D. 1033 17/06/2022, CN00000023).

\bibliographystyle{IEEEtranS}
\bibliography{main}

\end{document}